\documentclass[authoryear,3p,11pt]{elsarticle}
\usepackage{amsmath,amssymb,mathtools,mathrsfs}

\usepackage{ulem}
\usepackage{cancel}
\usepackage{float}

\usepackage[pdftex, pdftitle={Article}, pdfauthor={Author}]{hyperref} 
\usepackage[usenames,dvipsnames]{xcolor}
\usepackage{booktabs}
\usepackage{multirow}

\newcommand{\xiStar}{\xi_\ast}
\newcommand{\upd}{\mathrm{d}}
\newcommand{\tDelta}{\tilde{\Delta}}

\renewcommand{\H}{H_0}
\newcommand{\V}{V_0}
\newcommand{\tV}{\tilde{V}}
\newcommand{\tH}{\tilde{H}}
\newcommand{\zast}{z_\ast}
\newcommand{\xast}{x_\ast}

\usepackage[symbol]{footmisc}

\journal{Elsevier}
\usepackage[nodayofweek,level]{datetime}
\newdate{date}{15}{06}{2022}
\date{\displaydate{date}}

\begin{document}

\begin{frontmatter}


\title{Shape-morphing structures based on perforated kirigami}


\author{Yunlan Zhang$^{1,4}$, Jingyi Yang$^{1}$, Mingchao Liu$^{2,}$\footnote[1]{\textit{Corresponding Author: mingchao.liu@ntu.edu.sg (M.L.)}} and Dominic Vella$^{{3}}$}

\address{$^{1}$\:Department of Engineering Science, University of Oxford, Parks Rd, Oxford, OX1 3JP, UK\\
$^{2}$\:School of Mechanical and Aerospace Engineering, Nanyang Technological University, Singapore 639798, Republic of Singapore\\
$^{3}$\:Mathematical Institute, University of Oxford, Woodstock Rd, Oxford, OX2 6GG, UK\\
$^{4}$\:Department of Civil, Architectural and Environmental Engineering, University of Texas at Austin, Austin, TX 78712, United States of America
}


\begin{abstract}
Shape-morphing structures, which are able to change their shapes from one state to another, are important in a wide range of engineering applications. A popular scenario is morphing from an initial two-dimensional (2D) shape that is flat to a three-dimensional (3D) target shape. One of the exciting manufacturing paradigms is transforming flat 2D sheets with prescribed cuts (i.e. kirigami) into 3D structures. By employing the formalism of the ‘tapered elastica’ equation, we develop an inverse design framework to predict the shape of the 2D cut pattern that would generate a desired axisymmetric 3D shape. Our previous work has shown that tessellated 3D structures can be achieved by designing both the width and thickness of the cut 2D sheet to have particular tapered designs. However, the fabrication of a sample with variable thickness is quite challenging. Here we propose a new strategy -- perforating the cut sheet with tapered width but uniform thickness to introduce a distribution of porosity. We refer to this strategy as perforated kirigami and show how the porosity function can be calculated from our theoretical model. The porosity distribution can easily be realized by laser cutting and modifies the bending stiffness of the sheet to yield a desired elastic deformation upon buckling. To verify our theoretical approach, we conduct finite element simulations and physical experiments. We also examine the loading-bearing capacity of morphed structures via indentation tests in both FEM simulations and experiments. As an example, the relationship between the measured geometric rigidity of morphed half-ellipsoids and their aspect ratio is investigated in details.

\end{abstract}


\end{frontmatter}

\section{Introduction}
\label{sec:Introduction}

Structures that can adapt to different environments and complete various tasks by changing shapes from one state to another are referred to as shape-morphing structures \cite[]{oliver2016morphing}. Shape-morphing structures exist extensively as biological organisms in nature \cite[]{huang2018differential}. They are also an emerging class of advanced structures with engineering applications, including flexible electronics \cite[]{wang2019shape}, microfluidics \cite[]{wang2022shape}, energy devices \cite[]{zhao2021shape}, soft robotics \cite[]{lum2016shape,liu2021robotic}, deployable space structures \cite[]{mccue2021controlled} and aircraft drag control systems \cite[]{sofla2010shape,li2018review}, etc. Many mechanisms have been explored to achieve shape morphing in recent years, including pneumatic inflation \cite[]{siefert2019bio}, thermal expansion \cite[]{boley2019shape,zou2022encoding}, chemical swelling \cite[]{nojoomi2018bioinspired} and solvent diffusion \cite[]{yiming2020mechanics,tao2021morphing}, as well as magnetic \cite[]{nguyen2012morphing,bastola2021shape,tang2021programmable} and mechanical loading \cite[]{liu2020tapered,meng2022deployable}.

Among many engineering applications, morphing from a flat two-dimensional (2D) sheet into a three-dimensional (3D) structure with particular curvature distribution is one of the most common scenarios and has attracted much attention \cite[]{van2018programming,nojoomi20212d}. However, a significant challenge needs to be conquered first: to prevent the geometrical incompatibility between curved and flat surfaces that Gauss’ \emph{Theorema Egregium} depicts, i.e.~that deformations that preserve length or area (isometries) cannot change the Gaussian curvature of a surface \cite[]{gauss1828disquisitiones}. Creating localized stretching or expanding materials is one possible strategy. It can be achieved by either applying multifarious external physical effects (pressure, heat, and light) to actuate the local expansion or by using chemical stimuli to the material with elaborate micro-structures \cite[]{nojoomi2018bioinspired,siefert2019bio,boley2019shape,liu2021frustrating}. Nevertheless, since thin materials resist changes of length significantly, these approaches all depend on unconventional responsive materials and/or complex manufacturing techniques \cite[]{piedade20194d} to generate high stress/strain concentration at the areas that stretch locally \cite[]{vella2019buffering}.

Instead of locally stretching materials, another strategy is to buffer the geometric incompatibility by removing materials locally, i.e.~making cuts at specific positions in a material, known as kirigami \cite[]{neville2016shape}. By programming the cut pattern of kirigami, 3D shapes with different Gaussian curvatures can be generated \cite[]{callens2018flat,choi2019programming}. It should be noted that the curvature here referred to is ‘Apparent Gaussian Curvature’ (AGC), which means that, even though the 3D structure has distributed curvature at a global level, the local elements remain largely planar \cite[]{liu2020tapered}. This method has been broadly adopted for fabricating inorganic flexible electronic devices with complex 3D shapes, since it is easier to avoid significant strain \cite[]{zhang2017printing}. Using a design database of kirigami-based morphing structures and devices has become ubiquitous to achieve shape morphing from 2D patterns into 3D structures \cite[]{xu2015assembly}. On the contrary, how to design a kirigami 2D sheet that can morph into a given target profile had not been well-explored until analytical inverse design strategies were proposed recently \cite[]{liu2020tapered,fan2020inverse}. 

The inverse design framework developed by \citet{liu2020tapered} is based on the theory of the tapered elastica \cite[]{lee1993elastica}. More specifically, we considered an axisymmetric 3D structure as a target that can be formed from a flat 2D sheet with a tailored cut pattern by subjecting the sheet to a mechanical load. From a mechanics point of view, the 2D cut sheet can be viewed as multiple tapered elastic strips connected at one end to form the central hub. Since the tapered elastica theory can describe the deformation of each strip, we can therefore derive an explicit equation that correlates the shape of the 2D cut pattern (i.e. the width and thickness distributions of each elastic strip) and the curvature of the 3D target structure. We also demonstrated the feasibility of this inverse design framework by showing several typical examples.

Implementing the framework of \citet{liu2020tapered} to create tessellated morphing structures requires fabricating customized flat cut sheets with tapering of both the width and thickness. Although additive manufacturing technologies, such as 3D printing, allow the manufacture of such customized sheets \cite[]{isa2019five}, it is challenging to precisely control the thickness of a preformed flat sheet, especially for a sheet with small thickness and/or made of brittle materials \cite[]{adnan2017springback}. Looking into the underlying morphing mechanism, we can realize that the fundamental effect of changing the width or thickness of the sheet is to tune the local bending stiffness \cite[]{lee1993elastica}. Inspired by recent investigations on the bending behaviour of perforated sheets \cite[]{pezzulla2020deformation,shrimali2021remarkable},we now propose an alternative strategy to tailor the local bending stiffness -- introducing distributed pores. These can be easily introduced by, for example, laser cutting and micro-fabrication techniques \cite[]{pezzulla2020deformation} and modify the bending stiffness of the beam in a similar manner to thickness variations. Together with the geometrical constraint of tessellation \cite[]{liu2020tapered}, we present a new paradigm of inverse design of shape-morphing structures in this work and demonstrate it with a series of examples. We also move beyond the simple design problem considered by \citet{liu2020tapered} to consider the structural stability of the resulting shape: while a tesselated 3D shape may resemble a standard shell, we consider their load-bearing capacity. This is an important consideration since most shape-morphing structures will be required to withstand the action of external loads in applications from aircraft wing, to automobile structures and space or ground infrastructures \cite[]{sofla2010shape,daynes2013review,kormanikova2017shape}. 
We therefore examine the loading-bearing capacity of the shape-morphing structures generated using this `perforated kirigami' under indentation tests; in particular, by combining experiments and FEM simulations, we find that these structures have an unusual geometry-dependent rigidity in comparison to continuous shells.

\section{Theoretical Model}
\label{sec:TheoreticalModel}

Our previous work developed a theoretical framework based on the tapered elastica theory for the inverse design of shape-morphing problems \cite[]{liu2020tapered}. The key idea is to control its local mechanical properties (namely the bending stiffness) by varying the geometry (i.e.~width and/or thickness) of the flat sheet. Since changing the thickness of a thin sheet is generally not easy to achieve, in this work we modify it, achieving the same ultimate effect through variations of the porosity (or the pore volume fraction) combined with tailoring of the width.

\begin{figure}[ht]
    \centering
    \vspace{0.5cm}
    \includegraphics[width=1.0\linewidth]{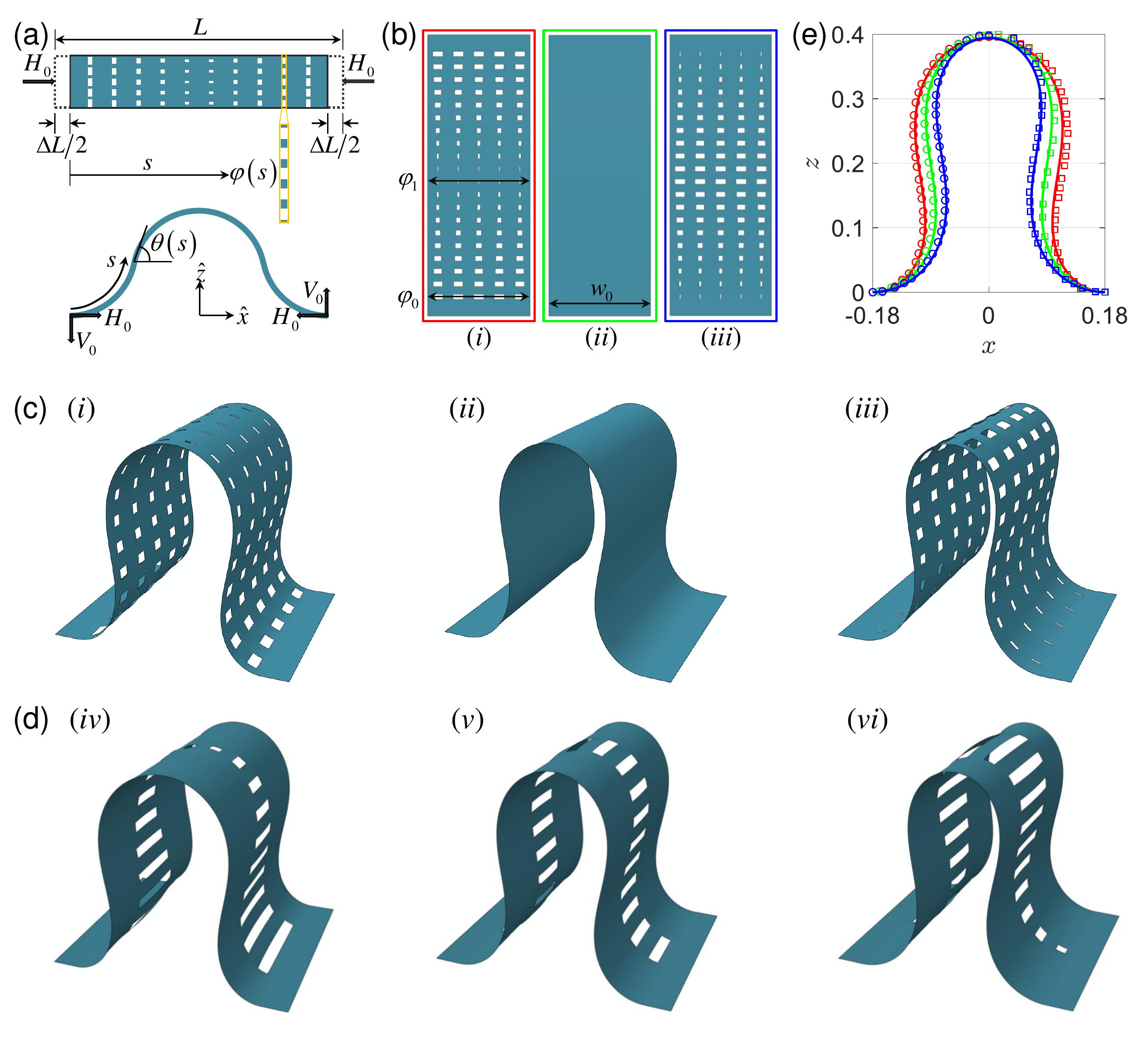}
    \caption{Porous elastic strips with distributed local porosity and their deformation subject to external load. (a) Schematic of a porous strip with distributed local porosity $\varphi(s)$ subject to a horizontal compressive load, $H_0$. The buckled shape is described by the intrinsic equation $\theta(s)$. (b) Representation of three strips with different local porosity distributions ($\varphi(s)$) but uniform width ($w_0$) and thickness ($t_0$); the local porosity ratio between the edge and centre varies in each case as follows: (\textit{i}) $\varphi_0/\varphi_1 = 10$, (\textit{ii}) $\varphi_0/\varphi_1 = 1$, (\textit{iii}) $\varphi_0/\varphi_1 = 0.1$. (c) The deformed shapes of those three porous strips displayed in (b). (d) The deformed shapes of three porous strips with same distributions of the local porosity corresponding to (c) but replacing a row of small pores by a single narrow slot.}
    \label{fig:fig1}
\end{figure}

To implement the inverse design strategy of shape-morphing by making cuts in a porous sheet, the first step is to correlate the bending stiffness of the porous structure and its micro-structure. The exact expression of the effective bending stiffness of a porous sheet is complex, and depends not only on the overall porosity but also on the spatial distribution and shape of the pores \cite[]{shrimali2021remarkable}. Here, to simplify the problem, we only focus on the local bending behaviour of the porous sheet. We define the pore volume fraction in a local area with width, $g$, where pores are located (see the zoom-in in fig.~\ref{fig:fig1}(a), marked as yellow) as the local porosity, $\varphi(s)$, which varies with the arc length, $s$. Following this definition, the local bending stiffness of a porous strip of length $L$ with non-uniform moment of inertia, $I(s)$, and distributed local porosity, $\varphi(s)$, can be written as
\begin{equation}
    B(s) = E_0 \cdot I(s) \cdot \left[ 1-\varphi(s) \right],
\label{eqn:1}
\end{equation} $E_0$ is the Young's modulus of the strip, which is a constant. (Here, we focus exclusively on changes to $I(s)$ achieved by varying the width, $w(s)$, keeping the thickness uniform, $t_0$; we therefore have $I\left( s \right)={w\left( s \right) \cdot {t_0^3}}/12$ but have left the general expression for completeness.)

Note that the `local porosity', $\varphi(s)$, in \eqref{eqn:1} is different from the global porosity, which we denote $\phi$ and measures the overall volume fraction of pores. In this way, it is possible to tune the global porosity by changing the width of the local area, $g$, whilst maintaining the same distribution of the local porosity. This advantage is important for customizing the load-bearing capacity of the morphed structures, which will be discussed further in Section \ref{sec:Rigidity}.

As shown in fig.~\ref{fig:fig1}(a), we consider an elastic porous strip subject to horizontal and vertical forces, $\H$ and $\V$, at its two ends, respectively. Recalling the tapered \textit{elastica} equation \cite[]{Stuart2000,liu2020tapered}, we follow \citet{liu2020tapered} to write the intrinsic equation for the shape, $\theta (s)$, as
\begin{equation}
    - \frac{\upd}{{\upd s}}\left[B(s) \cdot \frac{{\upd\theta (s) }}{{\upd s}} \right] = {w_0}\left[ {{\H}\sin \theta (s) + {\V}\cos \theta (s) } \right].
\label{eqn:2}
\end{equation} This is to be solved together with appropriate boundary conditions at both edges, e.g.~the clamped condition
\begin{equation}
    \theta \left( 0 \right) = \theta \left( L \right) = 0,
\label{eqn:3}
\end{equation} and the geometric constraints arising from the inextensibility, i.e. as shown in fig.~\ref{fig:fig1}(a), the beam accommodates a fixed amount of compression, $\Delta L$, which corresponds to
\begin{equation}
    \int_0^L\cos \theta(s)~\upd s = L - \Delta L = L(1 - \tDelta), 
\label{eqn:4}
\end{equation} where $\tDelta = \Delta L/L$. 

The complete shape of the buckled elastica, $[\hat{x}(s),\hat{z}(s)]$, may be determined from the intrinsic equation $\theta(s)$ by solving the geometrical relationships
\begin{equation}
    \frac{\upd \hat{x}}{\upd s}=\cos\theta,\quad \frac{\upd \hat{z}}{\upd s}=\sin\theta.
\label{eqn:5}
\end{equation}

Following \citet{liu2020tapered}, we non-dimensionalize the problem by letting ($\xi,x,z) = (s,\hat{x},\hat{z})/L$ and $\omega({\xi}) = w(s)/w_0$ (where $w_0$ is the value of the width at $s=0$); using eqs.~\eqref{eqn:1} and \eqref{eqn:5}, the tapered elastica equation \eqref{eqn:2} becomes
\begin{equation}
    \frac{\upd}{\upd\xi}\left\{\omega( \xi) \cdot \left[1-\varphi(\xi)\right] \cdot \frac{\upd\theta}{\upd\xi} \right\} = - \tH\frac{{\upd z}}{{\upd\xi }}-\tV\frac{\upd x}{\upd \xi},
\label{eqn:6}
\end{equation}
where $\tH = \left[(w_0 L^2)/(E_0I_0)\right] \cdot \H$ and $\tV = \left[(w_0 L^2)/(E_0I_0)\right] \cdot \V$ are the dimensionless forces and $I_0 = {w_0}{t_0^3}/12$. In addition, both the boundary condition \eqref{eqn:3} and the geometric constraint \eqref{eqn:4} are non-dimensionalized correspondingly.

Equation \eqref{eqn:6} may be referred to as the porous tapered elastica equation but it is worth noting that the main difference between this equation \eqref{eqn:6} and eq.~(8) of \cite{liu2020tapered} is simply the replacement of $T(\xi)^3$ by $\left[1-\varphi(\xi)\right]$: variations in thickness are replaced by variations of the local solid fraction. Before moving to present the inverse design framework, we first validate this variant of the tapered elastica equation.

According to eq.~\eqref{eqn:6}, we might expect to observe deformed shapes with different curvatures simply by controlling the distribution of the local porosity of strips. As a validation, we perform finite element method (FEM) simulations to examine the deformation of three elastic strips with different local porosity distributions but the same thickness and width, as shown in fig.~\ref{fig:fig1}(b) (from (\textit{i}) to (\textit{iii}) corresponding to $\varphi_0/\varphi_1 = 10$, 1 and 0.1), with clamped boundary conditions at both ends. The deformed shapes of three strips obtained from FEM simulations are shown in fig.~\ref{fig:fig1}(c). Given the difficulties of cutting small pores in a narrow strip, we also check the cases with the same local porosity distribution but replacing a row of small pores by a single narrow slot (with the same pore value ratio in the local area). The corresponding deformed shapes obtained from FEM simulations are given in fig.~\ref{fig:fig1}(d). The quantitative comparison of the deformed profiles between FEM simulations and theoretical predictions is presented in fig.~\ref{fig:fig1}(e), where the symbols on the left- and right-hand side correspond to the shapes in figs.~\ref{fig:fig1}(c) and (d), respectively.

As expected, controlling the distribution of the local porosity of 2D strips with uniform thickness and width leads to different deformed shapes upon buckling. More specifically, compare the shape of the uniform strip (either without pores, see fig.~\ref{fig:fig1}(c)-(\textit{ii}), or with uniform local porosity, see fig.~\ref{fig:fig1}(d)-(\textit{v}), with profiles represented in (e) by green curves), with a strip of larger local porosity at the edges, which exhibits a fatter deformed shape, see figs.~\ref{fig:fig1}(c)-(\textit{i}) and (d)-(\textit{iv}) and the red curve in (e); while the larger local porosity in the centre of the strip with lead to slimmer deformed shape, see figs.~\ref{fig:fig1}(c)-(\textit{iii}) and (d)-(\textit{vi}) and the blue curve in (e). According to eq.~\eqref{eqn:1}, it is natural to understand that the large local porosity will induce sharp curvature because the local bending stiffness is weakened when more material is removed locally. Now, we move on to consider how a desired 3D tessellated shape can be inversely designed by choosing an appropriate bending stiffness of each strip within the perforated kirigami, which in turn can be chosen by tapering the width and making pores to match the desired distribution of the local porosity.

\section{Inverse Design of Tessellated 3D Structures}
\label{sec:InverseDesign}

\subsection{Theoretical formulation}
\label{sec:InverseDesignTheory}

The primary goal of this work is to achieve the inverse design of tessellated 3D structures. As demonstrated in our previous work \cite[]{liu2020tapered}, we can form an axisymmetric morphable 3D structure by connecting several tapered elastic strips through a central hub. In this case, two tapering parameters within each strip should play the role simultaneously to satisfy both the geometric constraint condition of the tessellation and the local bending stiffness matching the desired curvature distribution. 

Following the procedure developed in \cite{liu2020tapered}, we determine the relationship between the geometric parameters of the 2D tapered porous sheet and the curvature of the 3D desired shape by integrating \eqref{eqn:6} once,
\begin{equation}
    \omega(\xi) \cdot \left[1-\varphi(\xi)\right] =\frac{\tH(z_\ast-z)+\tV(x_\ast-x)}{\upd \theta/\upd \xi}.
\label{eqn:7}
\end{equation} 

It is clearly seen that there are two tailoring variables, the width profile, $\omega(\xi)$, and the local porosity, $\varphi(\xi)$. Following \citet{liu2020tapered}, we determine the width tapering to satisfy the geometric constraint of tessellation: edges of neighbouring strips touch each other throughout. This means the width profile, $\omega(\xi)$, must be compatible with the deformation in that $\omega(\xi) = (2L/w_0)\cdot x(\xi)\cdot \tan(\pi/N)$, where $N$ is the number of strips we choose to make the morphing structure, and $x(\xi) =1 + \varepsilon-\tDelta - \int_{0}^\xi \cos \theta(\xi')~\upd \xi'$ is the x-coordinate of a particular element in the deformed configuration, in which, $\varepsilon$ is the radius of the central hub, as marked in fig.~\ref{fig:fig2}(a). We can therefore determine $\omega(\xi)$, which implicitly contains the local inclination angle, $\theta(\xi)$, as

\begin{equation}
    \omega\bigl[ \xi;\theta(\xi)\bigr]=\frac{2 \tan(\pi/N) \cdot L}{w_0}\left[\varepsilon + X(\xi) \right],
\label{eqn:8}
\end{equation}
where $X(\xi)=1 - \tDelta - \int_0^\xi\cos \theta(\xi')~\upd\xi'$. Note that this is determined independently of the local porosity; by substituting eq.~\eqref{eqn:8} into eq.~\eqref{eqn:6} (assuming $\varphi(\xi) = 0$) and applying particular boundary conditions, we achieve tessellating 3D axisymmetric shapes. However, the profile of these shapes are not those desired. To match the desired profile, whilst maintaining the tessellation, we tune the distribution of local porosity. Rearranging eq.~\eqref{eqn:7}, we relate the local porosity, $\varphi(\xi)$, to the curvature of the desired shape, $\theta(\xi)$, as

\begin{equation}
    \varphi(\xi) = 1-\frac{\tH(\zast - z)+\tV(\xast-x)}{\omega(\xi) \cdot \theta_\xi}.
\label{eqn:9}
\end{equation} 

For a given desired 3D target structure, once we extracted its curvature distribution, $\theta(\xi)$, as the input information, we can solve eqs.~\eqref{eqn:8} and \eqref{eqn:9}, together with appropriate boundary conditions, to obtain the geometric information of the cut pattern of the 2D flat sheet (both the width, $\omega(\xi)$, and the local porosity distribution, $\varphi(\xi)$).

Here it should be noted that, eq.~\eqref{eqn:8} can be directly solved once we have the curvature information from the target shape. However, to solve eq.~\eqref{eqn:9}, we need to determine the unknown parameters, $\H$ and $\zast$ (with horizontal load only) or $\V$ and $\xast$ (with vertical load only), or all of them together (with both horizontal and vertical loads simultaneously). We take the simplified situation in which the structure is subject only to a horizontal load as an example; then eq.~\eqref{eqn:9} can be reduced to

\begin{equation}
    \varphi(\xi) = 1-\frac{\tH(\zast - z)}{\omega(\xi) \cdot \theta_\xi}.
\label{eqn:10}
\end{equation} 

According to \cite{liu2020tapered}, we can classify two types of situations depending on whether the desired 3D shape has an inflection point, i.e.~whether there is any intermediate $\xiStar\in[0,1]$ satisfying $\theta_\xi(\xiStar)=0$: the two cases are denoted (a), no inflection point, and (b), one inflection point. We can summarize the solution of the two unknown parameters ($\H$ and $\zast$) corresponding to the particular boundary conditions, for situation (a)
\begin{equation}
\tH = \frac{\omega(0)\theta_\xi(0)-\omega(1)\theta_\xi(1)}{y(1)},\quad \zast = \frac{\omega(0)\theta_\xi(0)}{\tH},
\label{eqn:11}
\end{equation} 
and, for situation (b)
\begin{equation}
\zast = \int_0^{\xiStar} \sin \theta~\upd\xi,\quad \tH = \frac{\theta_\xi(0)}{\zast}.
\label{eqn:12}
\end{equation} 

Substituting eq.~\eqref{eqn:11} or \eqref{eqn:12} into eq.~\eqref{eqn:10}, together with eq.~\eqref{eqn:8}, the inverse design problem is fully solved. Similar results can be obtained for the situation of a vertical load only. It should also be noted that the local porosity must satisfy $\varphi(\xi)\in[0,1]$; to ensure this, we scale the local porosity calculated from eq.~\eqref{eqn:10} as $\bar{\varphi}(\xi) = \left[1-\varphi(\xi)\right]/\max\left[1-\varphi(\xi)\right]$.

\subsection{Demonstration of inverse design strategy}
\label{sec:InverseDesignDemonstration}

\begin{figure}[ht]
    \centering
    \vspace{0.5cm}
    \includegraphics[width=1.0\linewidth]{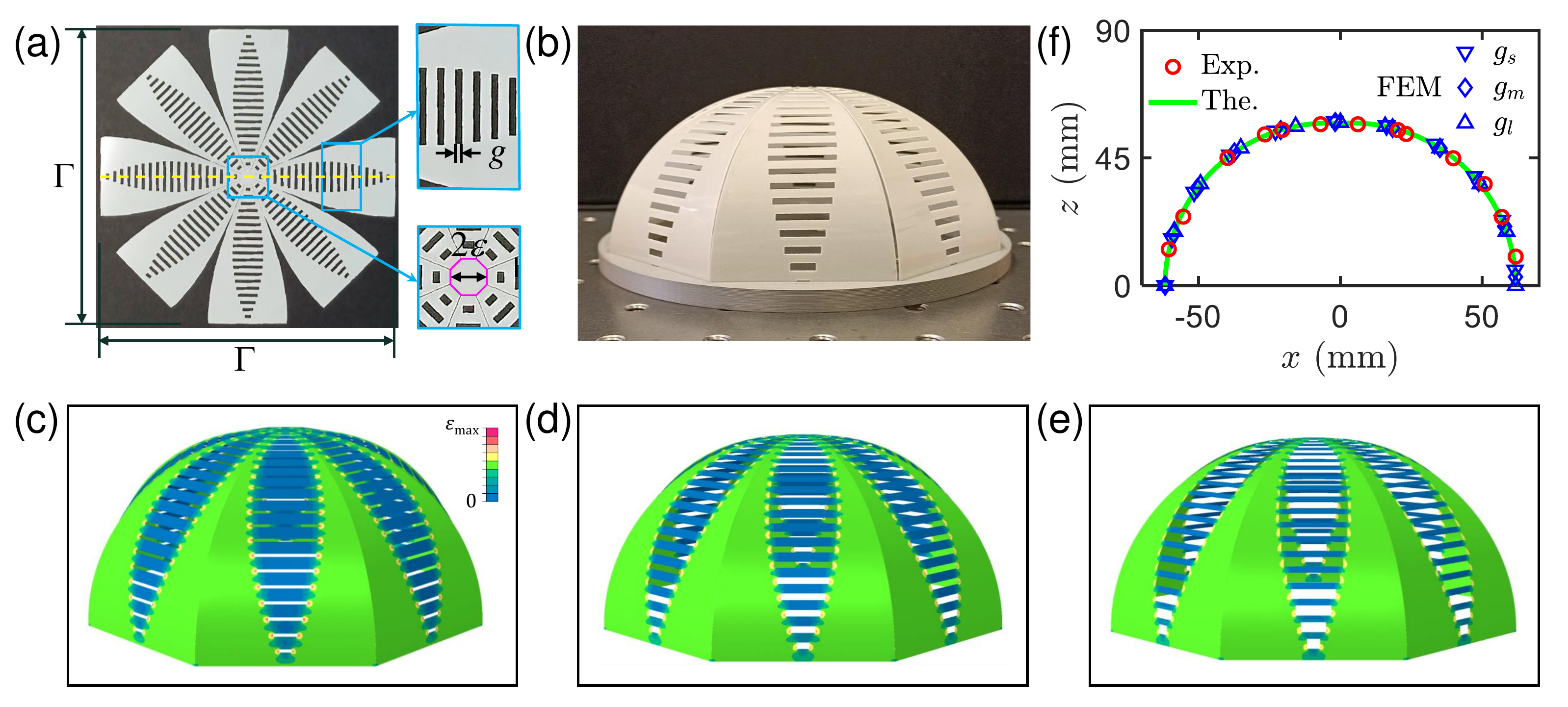}
    \caption{Inverse design demonstration of a morphing to a hemisphere in FE simulations and experiment. (a) The cut patterns of the 2D flat sheet predicted by theory to generate the 3D hemisphere. (b) A morphed hemisphere with $g_m = 1.0g$ (where $g/\Gamma = 0.0096$) obtained in physical experiment. (c-e) The 3D morphed hemispheres obtained in FEM simulations with dimensionless slot size: (c) $g_s = 0.5g$, (d) $g_s = 1.0g$ and (e) $g_s = 1.5g$. (f) Comparison of the profiles of the 3D morphed hemispheres obtained from the theory, FEM simulations and experiment.}
    \label{fig:fig2}
\end{figure}

As a first demonstration of the inverse design framework proposed in Section \ref{sec:InverseDesignTheory}, we choose a hemisphere as the target 3D structure. This structure can be morphed by applying horizontal loads to a 2D cut sheet with distributed porosity. We design the 2D cut pattern (the distribution of both width, $w(s)$, and local porosity, $\varphi(s)$) by harnessing the theoretical model; in particular, eqs.~\eqref{eqn:8}, \eqref{eqn:10} and \eqref{eqn:11} are used since the target shape is without inflection point. To simplify the fabrication, we consider a narrow slot with width, $g$, to represent a row of small pores, similar to the cases shown in fig.~\ref{fig:fig1}(d). Here we choose to use eight strips ($N=8$) to make up the structure. The designed 2D cut pattern is fabricated by laser cutting a uniform square flat sheet with dimensions $\Gamma \times \Gamma \times t$ (with $\Gamma = 189$ mm and $t = 0.3$ mm are the width and thickness of the flat sheet, respectively), and the slot width is set as $g/\Gamma = 0.0096$, see \ref{methods_fabrication} for more details. The cut pattern of the flat sheet is shown in fig.~\ref{fig:fig2}(a). 

To ensure the cut sheet is subject to precise boundary conditions at the distal edge of each strip, we also 3D printed a base with eight slots that match the width and thickness of the distal edge of each strip (see \ref{methods_fabrication} for more details). The position and inclination angle of each slot correspond to the boundary conditions of displacement and rotation angle required for the final shape. Inserting each strip into the slot tightly, we obtain the 3D morphed structure shown in fig.~\ref{fig:fig2}(b). We also perform FEM simulations to validate this example of morphing --- the morphed shape obtained from FEM is shown in fig.~\ref{fig:fig2}(d). The details of FEM simulation are given in \ref{methods_fem}. From both experiments and simulations, we see clearly that the axisymmetric tessellated 3D structure is obtained. 

We also simulate two cases with different slot widths in FEM simulations, comparing to those cases with $g_m = 1.0g$ shown in figs.~\ref{fig:fig2}(b) and (d), namely smaller slot, $g_s = 0.5g$, and larger slot, $g_l = 1.5g$; the obtained morphed shapes are shown in figs.~\ref{fig:fig2}(c) and (e), respectively. All these cases show the good match with the target shape; this indicates that the value of slot width $g$ doesn't affect the shape of the morphed structure. The color map of figs.~\ref{fig:fig2}(c-e) shows the distribution of strain within the morphed structure; the corresponding maximum values of the strain for (c-e) are $\epsilon_{max} < 0.5\%$. The extremely small strain confirms that the bending-induced deformation is almost isometric and indicate that this kind of design strategy can be easily implemented within a broad range of materials without inducing material failure.

Furthermore, the coordinates of the mid-line profile of the axisymmetric morphed 3D structures (along the yellow line as marked in fig.~\ref{fig:fig2}(a)) are extracted from both experiment (the shape in fig.~\ref{fig:fig2}(b)) and FEM simulations (figs.~\ref{fig:fig2}(c-e)). They are quantitatively compared with the profile predicted by the theoretical model (which is the same as the target shape, by construction) and shown in fig.~\ref{fig:fig2}(f). The excellent agreement between theory, experiment and simulations indicates that the inverse design method is effective. We therefore move on to explore the mechanical performance of the morphed 3D structures.

\section{Rigidity of Morphed 3D Structures}
\label{sec:Rigidity}

To create functional and resilient 3D structures via this kind of inverse design strategy, understanding the mechanical behaviour of the morphed structure is essential. Considering the created hemi-spherical morphing structure has the potential for application in engineering structures, such as infrastructures, we examine its rigidity in indentation as a proxy of its overall load-bearing capacity: how does geometry influence the structural rigidity within the linear elastic regime?

\begin{figure}[ht]
    \centering
    \vspace{0.5cm}
    \includegraphics[width=1.0\linewidth]{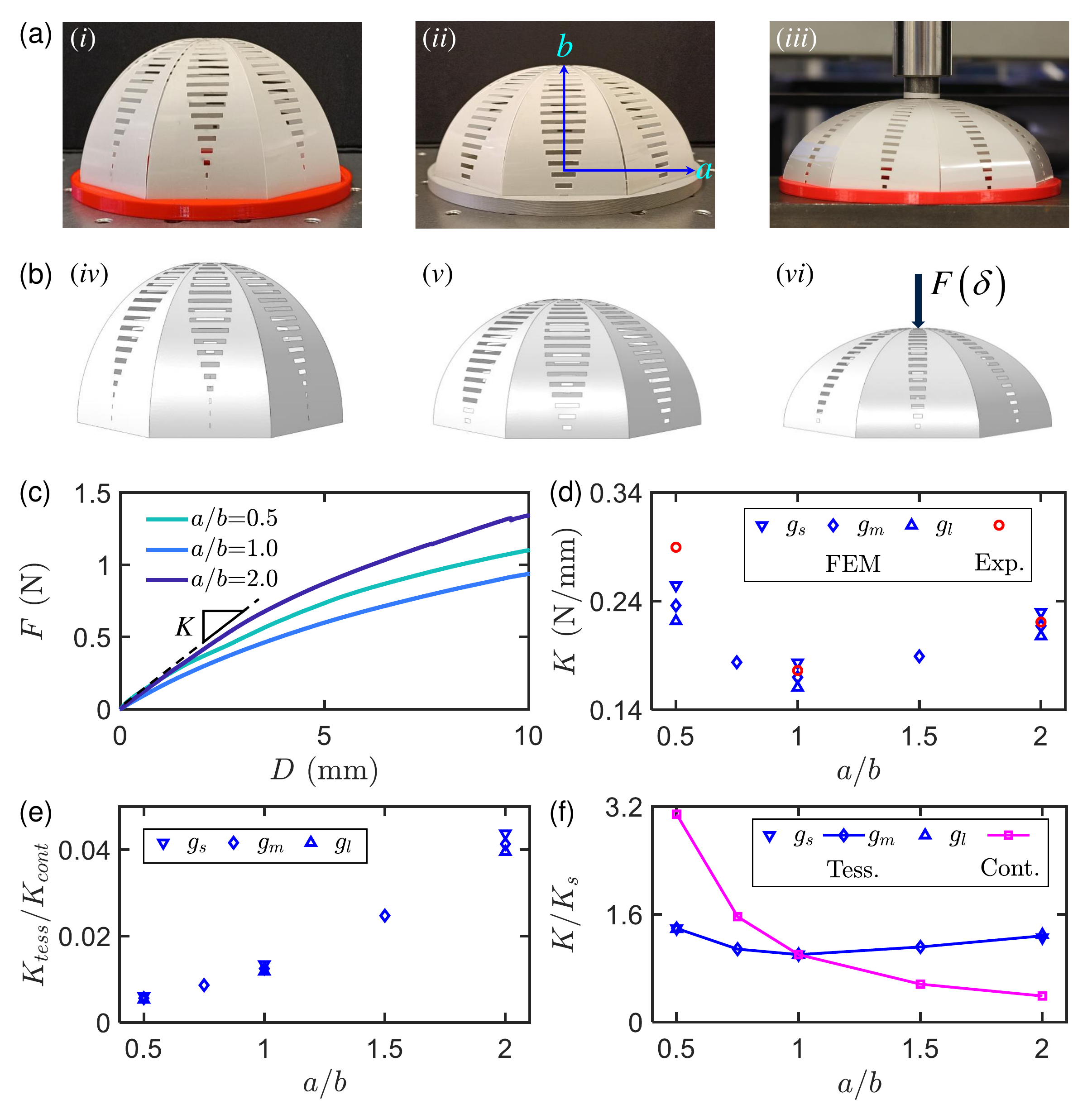}
    \caption{Experimental and numerical realization of half-ellipsoids with different aspect ratio and their rigidity measured in indentation tests. (a) The morphed half-ellipsoids obtained in physical experiments with (\textit{i}) $a/b =$ 0.5, (\textit{ii}) $a/b =$ 1.0 and (\textit{iii}) $a/b =$ 2.0. (b) The morphed half-ellipsoids obtained in FEM simulations, the shapes in (\textit{iv}-\textit{vi}) correspond to the physical models in (\textit{i}-\textit{iii}). (c) The force-displacement relationship of half-ellipsoids in indentation tests. (d) The rigidity of the half-ellipsoids measured from both experiments and FEM simulations. (e) The rigidity ratio between the morphed structures (referred to as the tessellated shells) and a continuous shell with the same aspect ratio. (f) The rigidity of both tessellated shells and continuous shells normalized by the results for the corresponding hemispheres (with $\textit{a/b}=1.0$).}
    \label{fig:fig3}
\end{figure}

Previous studies indicated that the stiffness of a continuous ellipsoidal shell is strongly dependent on its geometry \cite[]{vella2012indentation2,lazarus2012geometry}, in which the aspect ratio $a/b$ is an important geometric parameter. As an extension to the hemisphere (with $a/b = 1.0$), we also create several half-ellipsoid morphing structures with aspect ratio $a/b$ in the range $0.5\leq a/b\leq 2.0$. The experimental and numerical realization of three half-ellipsoids with $a/b$ = 0.5, 1.0 and 2.0 are shown in figs.~\ref{fig:fig3}(a)-(\textit{i}-\textit{iii}) and (b)-(\textit{iv}-\textit{vi}), respectively. Note that these structures are cut from square sheets with the same width, $\Gamma = 189$ mm. 

We quantify the structural rigidity of these morphed 3D structures via indentation tests. The test setup is displayed in fig.~\ref{fig:fig3}(a)-(\textit{iii}): The top central hub of the morphed structure is indented by a probe (the `indenter') with a flat tip, to a depth of $D = 10$ mm (see \ref{methods_indentation} for details). The measured relationships between the applied indentation depth, $D$, and the measured reaction force, $F$, are presented in fig.~\ref{fig:fig3}(c). The initial stiffness of each morphed structure is measured from the $F-D$ curves within the linear elastic regime ($D \lesssim t$, as suggested in \cite[]{lazarus2012geometry}); this defines the (linear) rigidity of the morphed structure. The measured value of $K$ is plotted as a function of $a/b$ in fig.~\ref{fig:fig3}(d) by red circles. We also implement the indentation test in FEM simulations for samples with $a/b$ = 0.5, 0.75, 1.0, 1.5, and 2.0, and the numerically measured rigidity are plotted in fig.~\ref{fig:fig3}(d) as blue diamonds; experiments and FEM simulations show consistent results here. Interestingly, the relationship between $K$ and $a/b$ is non-monotonic. This non-monotonic relationship is resulting from the complex interplay of geometric and structural factors but can be roughly correlated to the global porosity, $\phi$, which determines the weakening effect on the bending stiffness of strips by removing materials. We measure $\phi$ for morphed structures with different $a/b$ and find positive correlation between $K$ and $1-\phi$, which can be referred as the volume fraction of solid, with respect to $a/b$; that is, for given $a/b$, the morphed structure with smaller global porosity $\phi$ corresponding to larger rigidity $K$. 

As already discussed (Section 3.2), the same morphed shape can be realized by using cut patterns with the same distribution of local porosity and changing only the width of each slot. We therefore also perform FEM simulations to investigate the rigidity of these variants on a single morphed structure to quantify the role of slot width. More specifically, we evaluate the rigidity of structures with $a/b = 0.5$, 1 and 2 for both narrower and wider slots. The numerically measured values of the rigidity are also plotted in fig.~\ref{fig:fig3}(d). Comparing to the cases with $g_m$, and the same value of $a/b$, the narrower slot cases ($g_s=0.5g_m$) are stiffer while the wider slot cases ($g_l = 1.5g_m$) are softer. These findings are in line with expectations: in general, a morphed structure should be stiffer when we remove less material from the sheet with given cut pattern. However, this also provides a new opportunity to tune the rigidity of a given target structure.

The rigidity of continuous ellipsoidal shells is known to depend on the aspect ratio $a/b$ \cite[]{vella2012indentation2,lazarus2012geometry}. One natural question is that how stiff the morphed shell (the tessellated shell) is compared to the continuous shell of the same gross shape. \cite{vella2012indentation1} showed that the rigidity of a continuous shell with aspect ratio $a/b$ can be calculated as $K=8B/l_b^2$, where $l_b = \sqrt{(pR/Eh)}R$ is the bending length scale calculated with the radius of curvature local to the indentation point, $R=a^2/b$. We can therefore readily calculate the rigidity of continuous shells with different values of $a/b$ when the perimeter is fixed as a constant (which is constrained by the dimension of the flat sheet, $\Gamma$). The ratios between the rigidity of tessellated shells (with different slot width $g$) and continuous shells, $K_{tess}/K_{cont}$, are plotted as a function of $a/b$ in fig.~\ref{fig:fig3}(e). It is clearly seen that the ratio is increasing with the increase of $a/b$. As might be expected, the rigidity of tessellated shells are much smaller than the corresponding continuous shells in general; we find that $K_{tess}/K_{cont} < 0.05$: since the tessellated shell is comprised of eight separated strips, which are free to separate upon indentation, and therefore the constraint that is an integral part of the geometric rigidity of the continuous shell is easily released. In addition, to match the curvature distribution of the target structure, we introduced porosity into the cut sheet to tailor its bending stiffness (and hence shape), thereby further weakening the stiffness of the morphed structures.

To further explore the difference between the tessellated shell and the continuous shell, we normalize all the results of rigidity obtained from FEM simulations (and calculations) by their value for the equivalent hemisphere ($a/b=1.0$), i.e., $\bar{K} = K/K_s$ where $K_s = K(a/b=1.0)$, and show how the normalized rigidity $\bar{K}$ varies with the aspect ratio $a/b$ for both tessellated and continuous shells in fig.~\ref{fig:fig3}(f). There are two main important findings: First, the rigidity of tessellated shells with different slot width $g$ are collapsed to a single curve, which indicate that the relationship between $K$ and $a/b$ is only depending on the distribution of the local porosity; Second, different from the tessellated shells having a non-monotonic relationship between $\bar{K}$ and $a/b$ with minimal rigidity at $a/b = 1.0$, the continuous shells have monotonic decreasing $\bar{K}$ as a function of $a/b$. This difference of the dependence of $\bar{K}$ on $a/b$ can be naturally ascribed to the release of the geometric constraint due to the separation of the strips which form the tessellated shells, but also leave a detailed analysis on the deformation mechanism of the tessellated shells for a future study.

\section{Discussion and conclusion}
\label{sec:DiscussionConclusion}

\begin{figure}[ht]
    \centering
    \vspace{0.5cm}
    \includegraphics[width=0.9\linewidth]{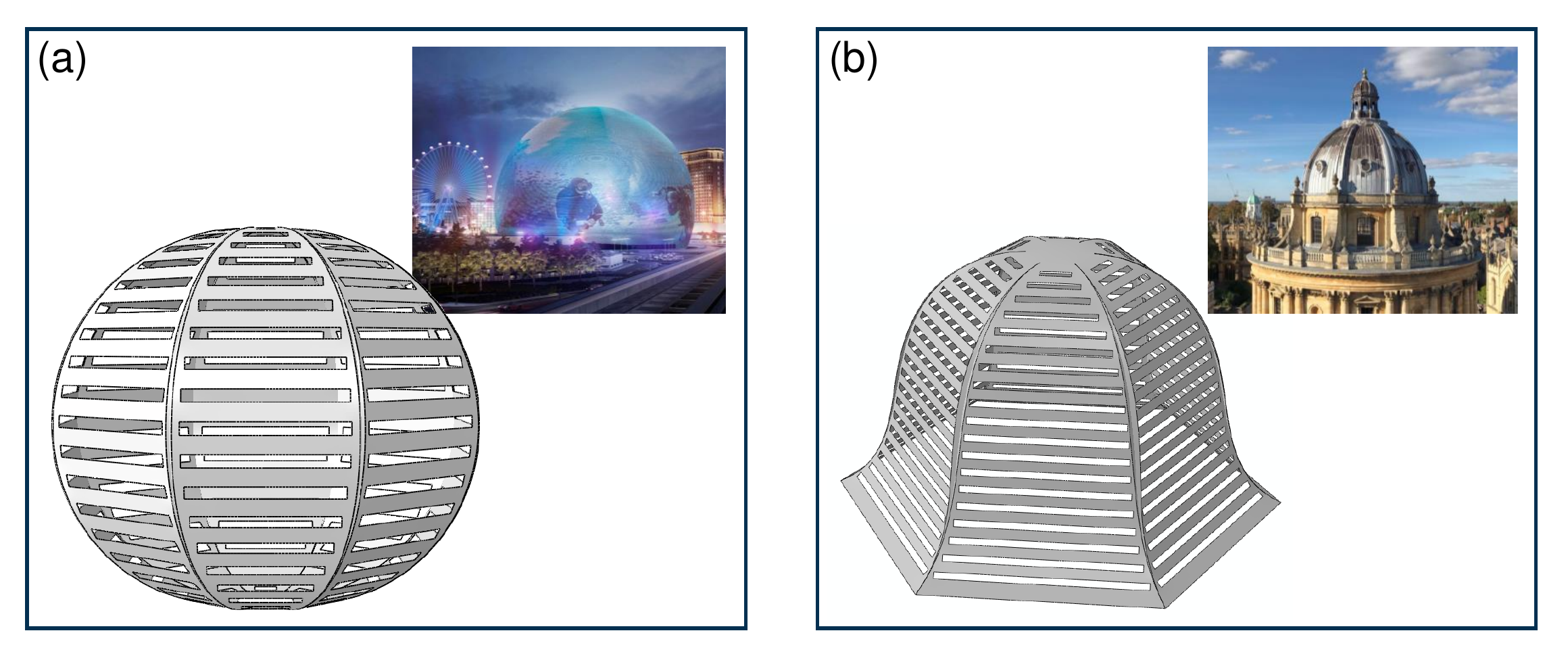}
    \caption{Additional examples of the shape-morphing structures with different shapes based on the inverse design framework. (a) A sphere with respect to the Madison Square Garden in Las Vegas (https://www.msgsphere.com/). (b) A roof structure with respect to the cupola of the Bodleian Library at the University of Oxford.
}
    \label{fig:fig4}
\end{figure}

In conclusion, this work introduced a general framework for the inverse design of shape-morphing structures based on kirigami sheets with distributed porosity. The tessellation pattern and the distribution of local porosity can be explicitly predicted by our theoretical model based on a modification of the tapered elastica equation. We validate this theoretical framework through a detailed demonstration of the inverse design of a hemispherical structure combined with physical experiments and FEM simulations. Noted that the design framework is general, regardless of the geometric shape and size of the target shape. We also presented two additional examples to demonstrate the generality of this method, as shown in fig. \ref{fig:fig4}. We can expect that this framework would be applicable for the design of large-span shell structures with potential applications in a wide range of engineering scenarios.

We also evaluated the load-bearing capacity of morphed half-ellipsoidal  structures with different aspect ratios via indentation tests; this revealed an interesting non-monotonic trend of the geometric rigidity with respect to the aspect ratio that has been confirmed by both simulations and experiments. We argued that this non-monotonic trend could be attributed to the interplay of geometric and structural factors and phenomenologically correlated to the global porosity. We hope this preliminary investigation on the mechanical responses of shape-morphing structures will provide guidance for the design of functional and resilient 3D structures via morphing strategy, and furthermore, to motivate more in-depth studies in this area.

\begin{appendix}
\setcounter{figure}{0}
\setcounter{equation}{0}
\setcounter{table}{0}
\renewcommand{\thefigure}{A.\arabic{figure}}
\renewcommand{\theequation}{A.\arabic{equation}}

\section{Experimental details}

\subsection{Fabrication}
\label{methods_fabrication}

The 2D sheet with designed pattern is cut from a Mylar film with dimensions of 189 mm $\times$ 189 mm $\times$ 0.30 mm using a 60 Walt Denford VLS 4.60 laser cutter with the power and speed settings listed in Table A.1. 
\begin{table}[h]
    \centering
	\caption{Parameters are used for laser cutting}
	\label{tab:parameters}
	\setlength{\tabcolsep}{15.2pt}
	\begin{tabular}{c l l l l}
		\toprule
		Parameters & Power (\%) & Speed (\%) & PPI (pulses per inch) & Z-axis (mm)\\
		\midrule
		Value & 56.1 & 100 & 500 & 0.3\\
		\bottomrule
	\end{tabular}
\end{table}

The cut sample is clamped to a solid base designed to impose the given boundary conditions, i.e. the displacement $\Delta$ and the inclined angle at the distal edge. The base is 3D printed in hard plastic, namely polylactic acid (PLA, with Young's modulus $\approx$ 2 GPa). Clamping of each strip is achieved by printing a slot within the base into which additional material at the end of each strip could be inserted, guaranteeing a particular edge inclination angle is achieved.

\subsection{Indentation test}
\label{methods_indentation}

We perform the indentation tests by using an cylindrical indenter made by aluminium alloy with a flat tip. The aluminium indenter (\textit{E} $\approx$ 69~GPa) has one magnitude higher stiffness than the Mylar (\textit{E} $\approx$ 1.9~GPa) sheet. The diameter of the indenter tip is 20 mm, which covers the entire central hub of the morphed structures. The indentation depth of $D = 10$ mm is applied at a rate of 1 mm/min (ensuring that the test is performed quasistatically). The applied force is recorded by an INSTRON testing machine with a 100 N load cell.

\section{FEM Simulations}
\label{methods_fem}
Computational analyses are performed by using the commercial finite element analysis software (ABAQUS 6.19) to simulate both the shape morphing process and the indentation test. 4-node shell elements (S4R) were used in every model. The material properties of Mylar are used in all the simulations (i.e.~Young's modulus \textit{E} = 1.9 GPa and Poisson’ s ratio \textit{v} = 0.3). There are two steps set in the FEM simulation: the first step, applying the appropriate boundary conditions (both the displacement and the inclined angle at the distal edge) to achieve the morphed state, and the coordinates of the profile of the morphed structures are measured; the second step, applying the vertical displacement to the central hub of the 3D morphed structure, and record the displacement and reaction force simultaneously.

\end{appendix}

\subsection*{Acknowledgments}
The research leading to these results has received funding from Nanyang Technological University via the Presidential Postdoctoral Fellowship (ML).

\bibliographystyle{elsarticle-num-names}
\bibliography{Bibliography.bib}

\end{document}